\documentclass[conference,a4paper]{IEEEtran}
\IEEEoverridecommandlockouts
\usepackage[nospace,noadjust]{cite}
\usepackage{amssymb,amsmath,amsthm,amsfonts,mathtools}
\usepackage{fontawesome}
\usepackage{mathabx}
\usepackage{algorithmic}
\usepackage{graphicx}
\usepackage{textcomp}
\usepackage{multirow}
\usepackage{xcolor}
\usepackage[inline]{enumitem}
\usepackage{url}
\usepackage[hidelinks]{hyperref} 
\usepackage{comment}
\usepackage{balance}
\usepackage{siunitx}
\usepackage{forest}
\usepackage{booktabs}
\usepackage{xspace}

\usepackage{subcaption}
\usepackage{svg}
\usepackage[belowskip=0pt,aboveskip=2pt]{caption}

\usepackage{pifont}

\def\BibTeX{{\rm B\kern-.05em{\sc i\kern-.025em b}\kern-.08em
    T\kern-.1667em\lower.7ex\hbox{E}\kern-.125emX}}
\begin{document}

\newcommand{\etal}{\emph{et~al.\xspace}}
\newcommand{\ie}{\emph{i.e.}, }
\newcommand{\eg}{\emph{e.g.}, }
\newcommand{\etc}{\emph{etc.\xspace}}
\newcommand{\vxv}[2]{$#1\!\times\!#2$}

\newcommand{\livesession}{$12,156$ }
\newcommand{\livetuple}{$521,138$ }
\newcommand{\liveCVrecords}{$507,290$ }
\newcommand{\VDvideoids}{$11,544$ }
\newcommand{\VDtuple}{$46,180$ }

\title{YTLive: A Dataset of Real-World YouTube Live Streaming Sessions
}

\author{
    \IEEEauthorblockN{
        Mojtaba Mozhganfar\IEEEauthorrefmark{1}, Pooya Jamshidi\IEEEauthorrefmark{1}, Seyyed Ali Aghamiri\IEEEauthorrefmark{1}, Mohsen Ghasemi\IEEEauthorrefmark{2}, \\Mahdi Dolati\IEEEauthorrefmark{2},
Farzad Tashtarian\IEEEauthorrefmark{3}, Ahmad Khonsari\IEEEauthorrefmark{1},
Christian Timmerer\IEEEauthorrefmark{3}
    }
    \vspace{1mm}
    
    \IEEEauthorblockA{
        \IEEEauthorrefmark{1}School of Electrical and Computer Engineering, College of Engineering, University of Tehran, Tehran, Iran\\    
        \IEEEauthorrefmark{2}Department of Computer Engineering, Sharif University of Technology, Tehran, Iran\\        
        \IEEEauthorrefmark{3}Christian Doppler Laboratory ATHENA, Alpen-Adria-Universität Klagenfurt, Austria\\
    }
}

\maketitle

\begin{abstract}
Live streaming plays a major role in today’s digital platforms, supporting entertainment, education, social media, etc. However, research in this field is limited by the lack of large, publicly available datasets that capture real-time viewer behavior at scale. To address this gap, we introduce YTLive, a public dataset focused on YouTube Live. Collected through the YouTube Researcher Program over May and June~2024, YTLive includes more than ~507{,}000 records from~12{,}156 live streams, tracking concurrent viewer counts at five-minute intervals along with precise broadcast durations. We describe the dataset design and collection process and present an initial analysis of temporal viewing patterns. Results show that viewer counts are higher and more stable on weekends, especially during afternoon hours. Shorter streams attract larger and more consistent audiences, while longer streams tend to grow slowly and exhibit greater variability. These insights have direct implications for adaptive streaming, resource allocation, and Quality of Experience (QoE) modeling. YTLive offers a timely, open resource to support reproducible research and system-level innovation in live streaming. The dataset is publicly available at: https://github.com/ghalandar/YTLive.
\end{abstract}


\section{Introduction}\label{sec:intro}
Live streaming has become an essential part of various applications, including entertainment, social media, telemedicine, and remote learning. As its use continues to grow, the demand for publicly available and high-quality datasets that capture real user engagement and live session characteristics over time to support research and development in this area is becoming increasingly important. According to~\cite{gvr}, the global live streaming market is projected to reach \$345.13 billion by 2030, highlighting the scale and potential impact of advancements in this field.

Existing datasets~\cite{10.1145/1298306.1298309,9148782,loh2022youtube,JIA2018112,8901772} often focused on on-demand video~(VoD) streaming or offline scenarios, offering limited insights into the distinct challenges posed by live streaming environments, including fluctuating audience sizes, variable broadcast\footnote{In this context, broadcast represents a live event on YouTube that has ended.} durations, and the real-time nature of viewer engagement. This gap in the literature restricts the development and evaluation of advanced algorithms for adaptive bitrate selection, Quality of Experience (QoE) modeling, and network optimization tailored specifically to live streaming. 

To address this limitation, the present study introduces a novel dataset collected over two months during May and June 2024, through participation in the YouTube Researcher Program~\cite{yrp}. By leveraging the YouTube Data API~\cite{yda}, we systematically gathered over half a million records capturing key aspects of live streaming sessions. Specifically, the dataset records the number of concurrent users watching YouTube live videos at 5-minute intervals and the duration of each live video, computed by observing and recording the start and end times of every broadcast. This large-scale, temporally granular dataset enables researchers to explore important questions such as audience fluctuation patterns, broadcast longevity, and their implications for system design and QoE modeling.
The main objectives of this study are threefold:
\begin{itemize}[leftmargin=*]
    \item To provide the research community with a comprehensive, openly accessible dataset reflecting real-world live streaming usage on a major global platform.
    \item To examine and characterize temporal viewing patterns and broadcast durations.
    \item To illustrate how this data can support broader research in adaptive streaming algorithms, network resource allocation, and user engagement analysis.
\end{itemize}

\begin{table}[t]
  \centering
  \caption{\small{Comparative overview of live streaming platforms datasets.}}
  \label{tab:rw}
  \renewcommand{\arraystretch}{1.3}
  \setlength{\tabcolsep}{1.5pt}
  \scriptsize
  \resizebox{\columnwidth}{!}{ 
  \begin{tabular}{ccccc}
    \hline
    \textbf{Study} & \textbf{Year}  & \textbf{Platform} & \textbf{Collection Period (Days)} & \textbf{Dataset availability}\\
    \hline
    Kaytoue \etal~\cite{10.1145/2187980.2188259} & 2012 & Twitch.tv & 102 & \cite{kaytouedataset}\\
    \hline
    Pires \etal~\cite{10.1145/2713168.2713195} & 2015 & Twitch \& YouTube live & 90 & \cite{youtubetwitchdataset}\\
    \hline
    Stohr \etal~\cite{10.1109/LCNW.2015.7365913} & 2015 & YouNow & 36 & Not available \\
    \hline
    Baccour \etal~\cite{10.1109/ICIoT48696.2020.9089607} & 2020 & Facebook & 33 & \cite{FacebookVideosLive18Dataset}\\
    \hline \hline
    YTLive (this study) & 2025 & YouTube live & 60 & \cite{YTLivedataset}\\
    \hline
  \end{tabular}
  }
  \vspace{-5mm}
\end{table}

\section{Related Work}\label{sec:rw}
User-generated live streaming is steadily growing, and optimizing these systems and advancing research require reliable data. In this section, we review existing datasets and experimental studies on user-generated live streaming and highlight the need for updated measurement data. Most important datasets are summarized in Table~\ref{tab:rw}.

\subsection{Platform-Scale Characterization Studies}
Early studies, such as those by Kaytoue \etal\ in 2012~\cite{10.1145/2187980.2188259} and Pires and Simon in 2015~\cite{10.1145/2713168.2713195}, provided important initial insights into user-generated live streaming. Building on these efforts, later research focused on newer and more diverse platforms. One major direction was the study of live streaming services integrated with social networks. For example, the FacebookVideoLive18 dataset offered the first large-scale analysis of Facebook Live, including user geo-locations~\cite{10.1109/ICIoT48696.2020.9089607}. Other works explored mobile-centric platforms like YouNow, highlighting their shorter session durations and unique quality-of-experience constraints~\cite{10.1109/LCNW.2015.7365913}.

Our work contributes a timely update to this body of research by introducing a large-scale dataset of YouTube Live, one of the largest and most influential platforms. The dataset was collected with fine-grained temporal resolution, enabling the analysis of current viewership patterns.

\subsection{System and Performance Optimization Studies}
Beyond platform-scale characterization, another major research stream has focused on optimizing the technical systems that support live streaming, especially with respect to QoE~\cite{10.1145/3304109.3306220,madanapalli2021modelinglivevideostreaming,10.1145/3394171.3413539}. These studies often target specific stages of the live delivery pipeline, such as content ingest~\cite{10.1145/3458305.3463375}, encoding~\cite{10.1109/QoMEX.2017.7965686}, and server-side delivery~\cite{10.1007/978-3-319-54328-4_5}.

Several works explore system architecture and performance. Deng \etal~\cite{10.1007/978-3-319-54328-4_5} conducted an in-depth analysis of Twitch's infrastructure, while Zhu \etal~\cite{10.1145/3458305.3463375} developed \textit{Livelyzer} to analyze ingest performance. At the network level, Madanapalli \etal~\cite{madanapalli2021modelinglivevideostreaming} developed machine learning models to distinguish live from VoD streams in real-time and infer QoE metrics from traffic traces.
Barman \etal~\cite{10.1109/QoMEX.2017.7965686} evaluated codec performance for live gaming, noting trade-offs between compression efficiency and computational overhead. Several studies, such as~\cite{10.1145/3304109.3306220,10.1145/3394171.3413539}, have turned to immersive formats such as 360° live video, identifying latency challenges tied to both camera and server components.

While these studies offer deep technical insights, many
rely on limited-access or proprietary data. Our publicly avail-
able dataset complements this work by providing large-scale viewership traces, which can directly inform
system-level decisions such as adaptive bitrate algorithms,
ingest server provisioning, and content replication strategies.

\subsection{Specific Data and User Behavior Studies}
A third direction of studies has focused on collecting and analyzing specialized datasets to support behavior modeling and application-specific tasks such as recommendation and dialogue generation. 
For example, Rappaz \etal~\cite{10.1145/3460231.3474267} introduced \textit{LiveRec}, a large-scale dataset of Twitch interactions designed to address the challenge of dynamic content availability.

Other work targets user interaction via multimodal signals: ($i$) Text Dialogue: \textit{LiveChat} captures personalized viewer-chat interactions from Chinese live streams~\cite{gao2023livechatlargescalepersonalizeddialogue}. ($ii$) Multimodal Intent: \textit{MMLSCU} supports cross-domain intent classification for stream comments~\cite{10.1145/3589334.3645677}. And ($iii$) Game-Scene Alignment: \textit{CS-lol} aligns live chat with in-game events for contextual understanding~\cite{10.1145/3576840.3578334}.
Additional datasets explore phenomena such as virtual gifting~\cite{kim2025spreadvirtualgiftinglive}, sales prediction in live e-commerce~\cite{XU2024114104}, and anomaly detection~\cite{10.1109/ICDE60146.2024.00362}. While these datasets enable fine-grained behavioral modeling, they often lack a broader context of platform-wide dynamics. Our dataset complements this trend by offering a macro-level view of audience behavior, suitable both as a standalone benchmark and as a contextual resource for downstream learning tasks.

In summary, prior work reveals a clear research trajectory—from foundational platform-scale studies to system-level optimization and specialized behavioral analyses. Yet, there remains a lack of publicly available datasets that capture the current scale and dynamics of platforms like YouTube Live. To support renewed investigation into this evolving ecosystem, we introduce a new dataset collected from YouTube Live, comprising 507{,}290 live sessions sampled at 5-minute intervals over two consecutive months in 2024. With its fine-grained temporal resolution and industry-scale scope, this dataset offers a valuable resource for examining modern live streaming behaviors and enables the research community to revisit long-standing assumptions or explore emerging questions in large-scale user analysis.

\section{Methodology and Dataset Structure}\label{sec:ds}
\textbf{Methodology}.
To construct and maintain an up-to-date dataset of YouTube live sessions, we developed an automated system consisting of three \textit{Python} scripts executed periodically using \textit{cron jobs} on an Amazon Elastic Compute Cloud (EC2) \texttt{t3.micro} instance. These scripts work in coordination to collect, filter, and enrich data related to active YouTube live streams. The system is designed to run continuously until the allocated YouTube Data API quota is fully utilized each day.

\begin{figure*}[t]
    \centering
    \includegraphics[width=\linewidth]{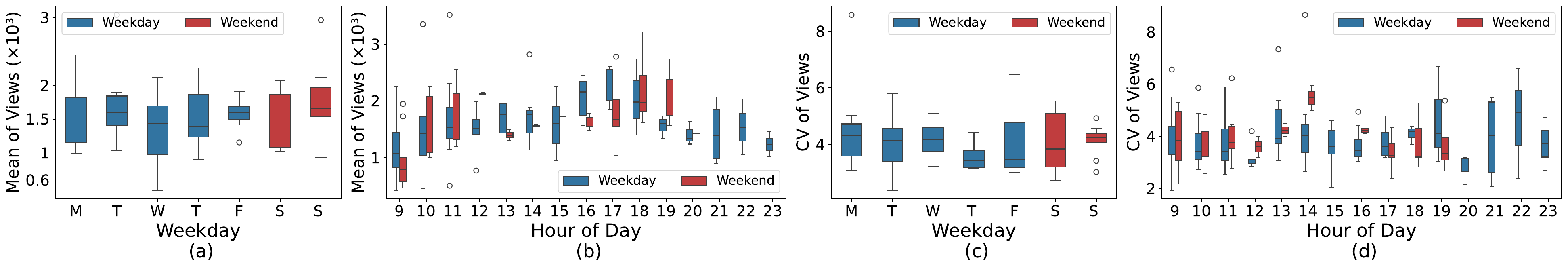}
    \caption{\small{Daily and hourly box plots of mean and coefficient of variation of viewership: (a) Viewership by day of week, (b) viewership by hour of day, (c) coefficient of variation of viewership by day of week, and (d) coefficient of variation by hour of day.}}
    \label{fig:results2}
    \label{fig:bp:stats}
    \vspace{-5mm}
\end{figure*}

\begin{figure*}[t]
  \centering
  \begin{subfigure}[b]{0.4\linewidth}
    \centering
    \includegraphics[width=\linewidth]{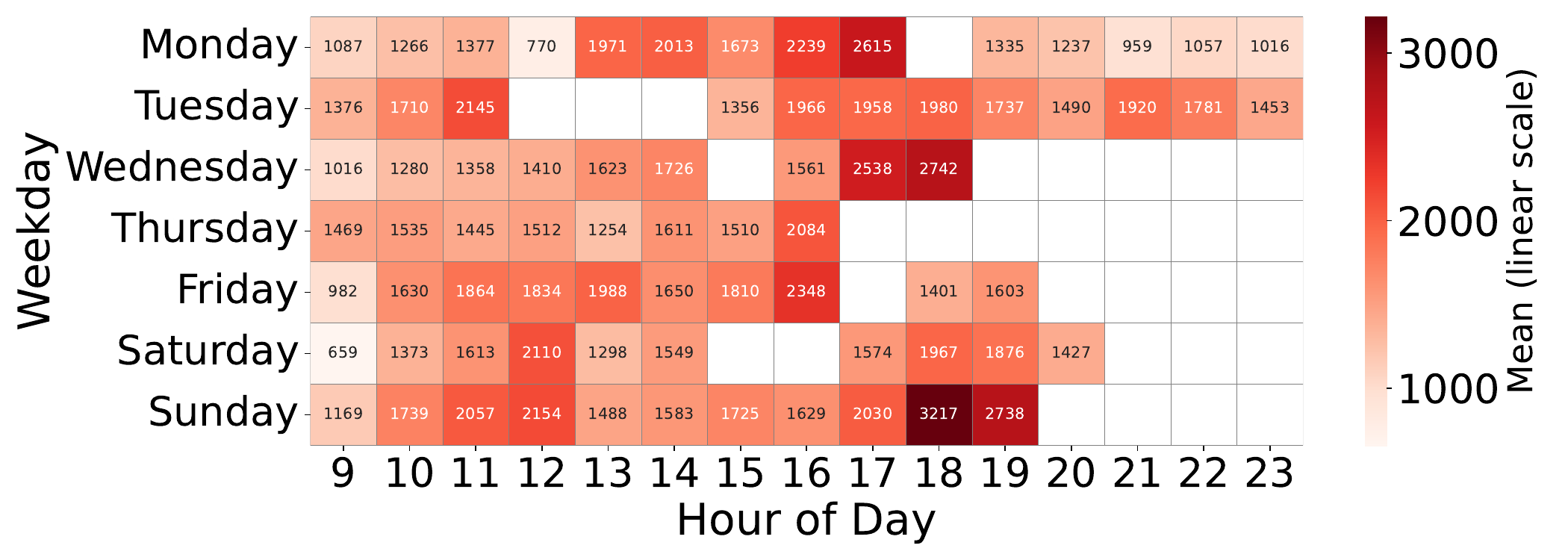}
    \caption{Mean hourly viewership by day of week.}
    \label{fig:heatmaps:mean}
  \end{subfigure}
  \hfil
  \begin{subfigure}[b]{0.45\linewidth}
    \centering
    \includegraphics[width=0.88\linewidth]{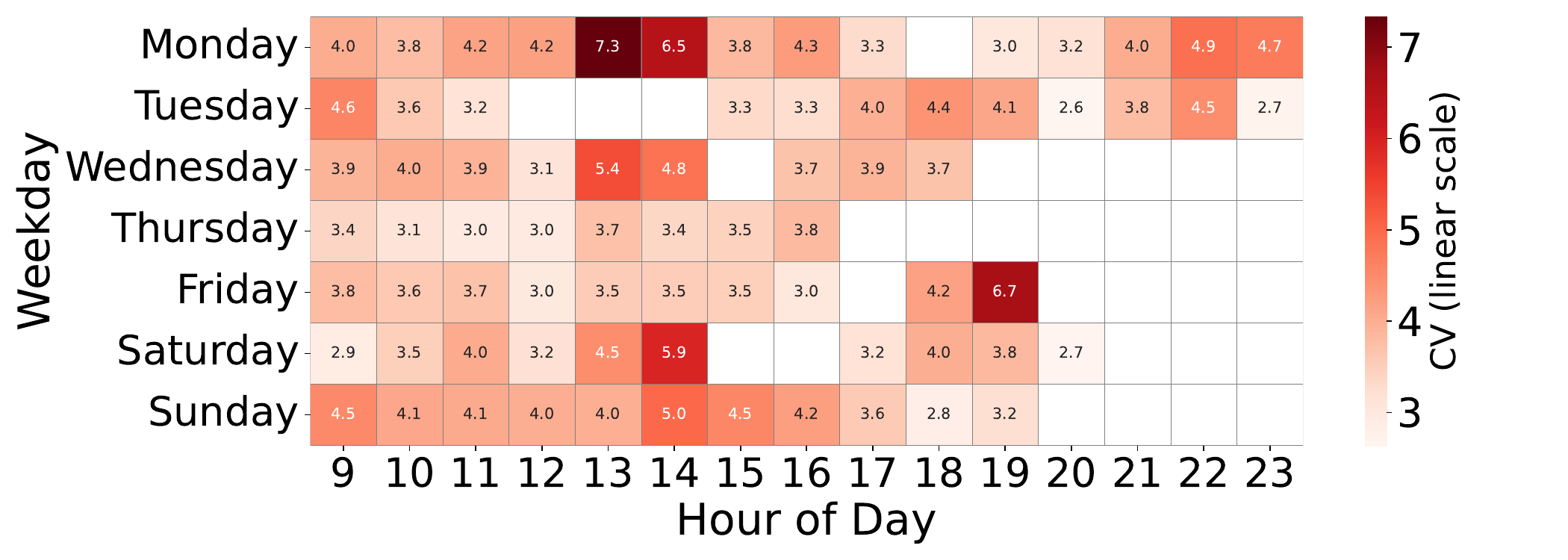}
    \caption{Coefficient of variation of hourly viewership by day of week.}
    \label{fig:heatmaps:std-mean}
  \end{subfigure}
  \caption{\small{Heatmaps of mean and coefficient of variation of viewership across the week.}}
  \vspace{-5mm}
  \label{fig:heatmaps}
\end{figure*}

The data collection system gather and refine metadata on active YouTube live streams. The first script runs hourly to discover ongoing live sessions using the YouTube Data API~\cite{yda} and appends the results to a shared \texttt{JSON} file. Immediately afterward, the second script executes to remove duplicate entries based on \texttt{videoId}, ensuring uniqueness in the dataset. Every five minutes, the third script enriches the data by fetching additional video details and track concurrent viewers across active live streams, while also capturing the precise start and end times of broadcasts to calculate their duration in minutes. All execution steps are logged separately for monitoring and debugging.
This modular and automated setup ensures timely collection of live video data, efficiently utilizing the API quota through scheduled intervals. The separation of discovery, cleaning, and enrichment tasks improves maintainability, while duplicate removal and stream-status filtering enhance data quality. Frequent updates allow the system to capture transient live sessions shortly after they start, resulting in a comprehensive and reliable dataset suitable for large-scale analysis of YouTube live sessions.

\textbf{Dataset Structure}. The YTLive dataset is organized into three main files: a \texttt{README.md} that provides an overview, and two CSV files, \texttt{durations.csv} and \texttt{viewers.csv}, which contain the measurement data.

The \texttt{durations.csv} comprises four columns detailing the total duration (in minutes) of each live video stream, where each stream is represented by its \textit{SHA-256} hash, along with precise start and end timestamps recorded with second-level accuracy (see Table~\ref{fields}).

Conversely, the \texttt{viewers.csv} file contains periodic measurements of viewer counts per stream at five-minute intervals. The first column denotes the timestamps at which the YouTube API was queried, while subsequent columns correspond to distinct hashed \texttt{videoId} identifiers for each stream. Each cell within these columns reflects the concurrent viewer count at the respective timestamp for the associated \texttt{videoId}.

Notably, the dataset exhibits several anomalous observations. Among these, a subset of streams remained continuously live throughout the two-month data collection period, without interruption. These persistent live streams represent a significant deviation from typical broadcasting patterns and warrant further investigation.

\begin{table}[b]
    \centering
    \vspace{-4mm}
    \caption{\small{YTLive fields and description}}
    \fontsize{7}{8}\selectfont
    \renewcommand{\arraystretch}{1.3}
    \begin{tabular}
    {p{1.7cm}lp{3.3cm}} \toprule
    \textbf{File Name} & \textbf{Field} & \textbf{Description} \\ \midrule
    \multirow{4}{0cm}{durations.csv} & videoId (hashed) & The string ID of the stream \\ \cline{2-3}
    & actualStartTime & The start timestamp of the stream \\ \cline{2-3} 
    & actualEndTime & The end timestamp of the stream \\ \cline{2-3}
    & duartion\_in\_minutes & Duration of stream in minutes \\ \midrule
    \multirow{2}{0cm}{viewers.csv} & Timestamp & Occurrence time of the event \\ \cline{2-3}
    & [\texttt{videoId}]s & No. of concurrent viewers \\ \bottomrule
\end{tabular}
        \label{fields}
    \vspace{-6mm}
\end{table}

\section{Results and Analysis}\label{sec:results}
We analyze the \livesession YouTube live streams introduced in Section~\ref{sec:ds} to identify key patterns and characteristics. To quantify the average viewership and its fluctuations, we use two metrics: the mean number of views and the coefficient of variation (CV), defined as the standard deviation divided by the mean.

Figure~\ref{fig:bp:stats}a shows that viewership, with a median of about $1500$ views, does not strongly depend on the day of the week. There is only a slight increase in the median and quartiles on Sundays. In contrast, Figure~\ref{fig:bp:stats}b shows a clear dependency on the hour of the day. Viewership is noticeably lower at the beginning and end of the day, with a peak between 17:00 and 19:00. Figures~\ref{fig:bp:stats}c and~\ref{fig:bp:stats}d present the variability of viewership. In Figure~\ref{fig:bp:stats}c, the median CV is around $4$, meaning that for half of the streams, the standard deviation of viewership is about four times the mean. The spread of CV values is limited, with the third quartile below $5$ and most values under $6$. The hourly breakdown in Figure~\ref{fig:bp:stats}d shows a similar trend, with a median CV close to $4$ and most values below $6$. 

\begin{figure}[b]
  \centering
  \vspace{-6mm}
  \begin{subfigure}[b]{0.45\linewidth}
    \centering
    \includegraphics[width=\linewidth]{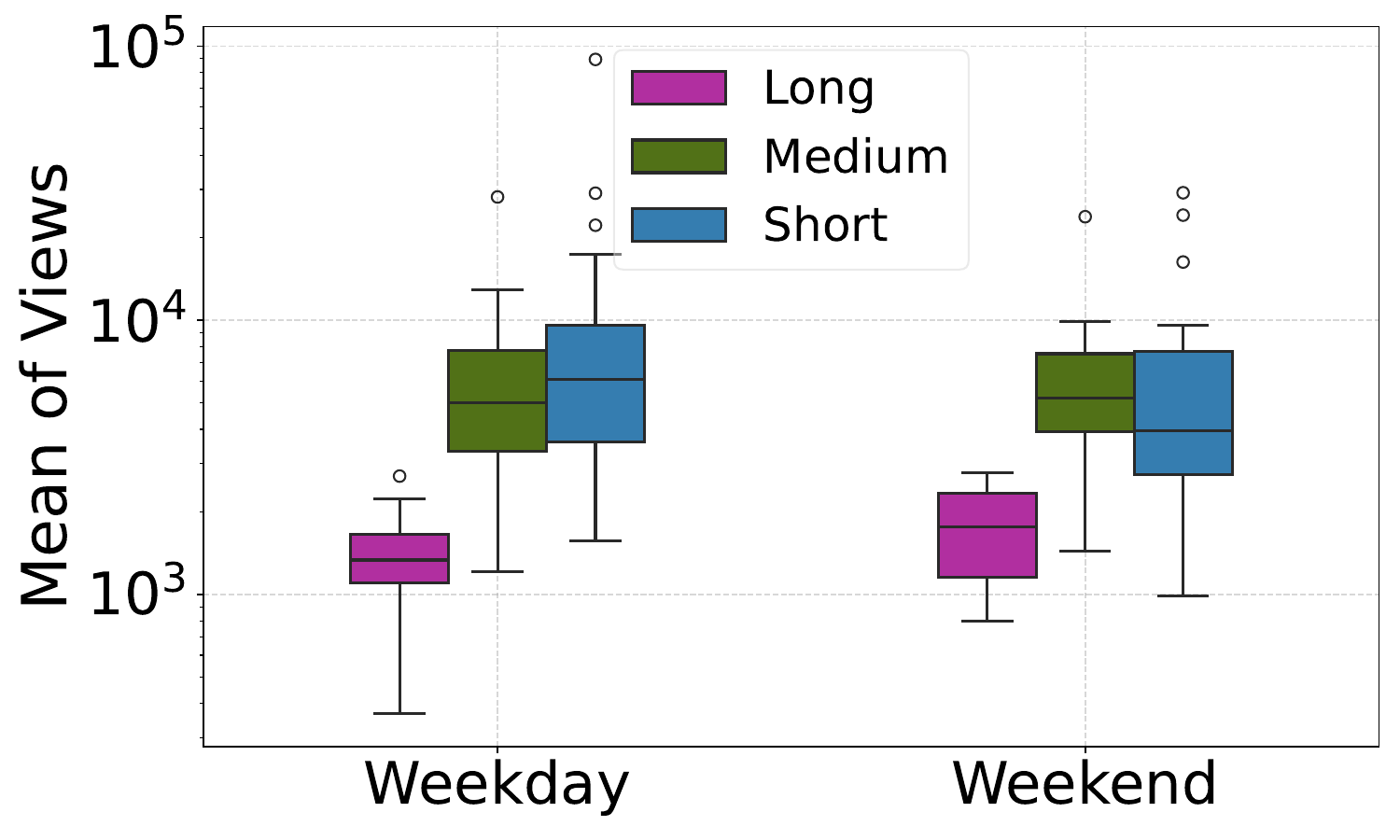}
    \caption{Mean of concurrent viewers}
    \label{fig:categorized:ubm}
  \end{subfigure}
  \hfill
  \begin{subfigure}[b]{0.45\linewidth}
    \centering
    \includegraphics[width=\linewidth]{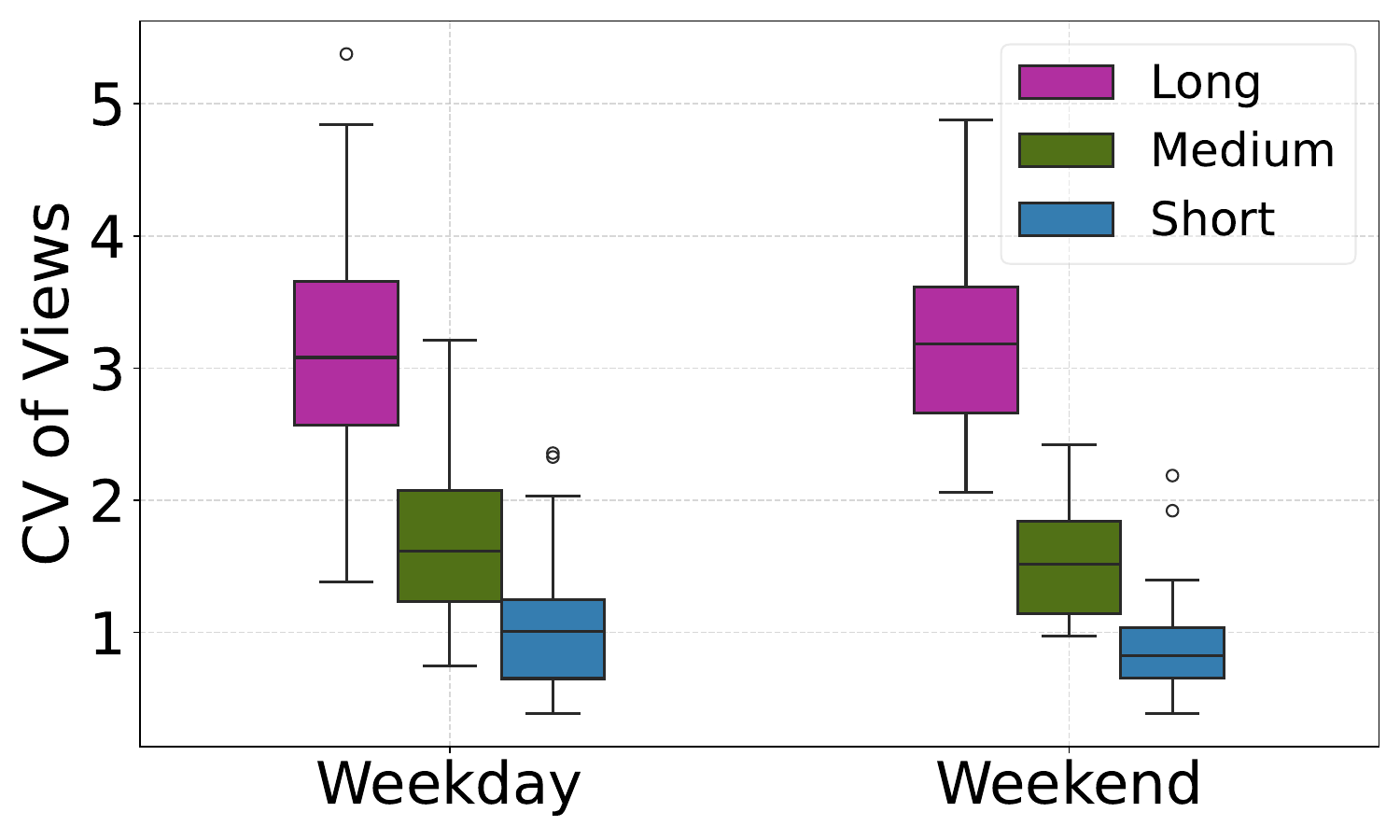}
    \caption{Coefficient of Variation}
    \label{fig:categorized:ubsom}
  \end{subfigure}
  \caption{\small{Viewer metrics by category: boxplots of mean and variability of concurrent viewers across video lengths and day types.}}
  \label{fig:categorized}
      \vspace{-6mm}
\end{figure}

We use two heatmaps to visualize viewership by hour and day of the week. Figure~\ref{fig:heatmaps:mean} shows the mean number of views. Sunday stands out with the highest viewership, especially between 11:00 and 19:00, peaking at 18:00. We also observe high viewership on Tuesday, Wednesday, and Friday afternoons. Early mornings and late nights generally have low viewership. Figure~\ref{fig:heatmaps:std-mean} shows the fluctuation of viewership, where darker colors indicate higher variability and lighter colors represent more stable traffic. For example, Sunday afternoons tend to have stable viewership, while Monday at 13:00 and Friday at 19:00 show higher variability. These patterns are useful for resource management. For busy and stable periods, such as Sunday evenings, we can allocate fixed resources. For times with high variability, it is better to use flexible systems that can scale up or down as needed. These insights can guide the efficient use of servers, bandwidth, and caching.

\begin{figure}[t]
    \centering
    \begin{minipage}{0.45\linewidth}
        \centering
        \includegraphics[width=\linewidth]{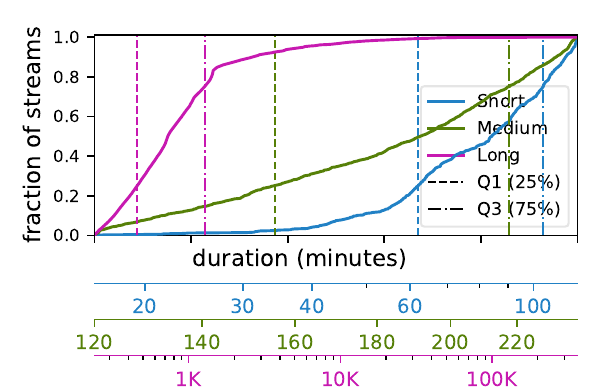}
        \caption{\small{CDF of video durations across categories}}
        \label{fig:categorized:cdfs}
    \end{minipage}
    \hfil
    \begin{minipage}{0.45\linewidth}
        \centering
        \includegraphics[width=\linewidth]{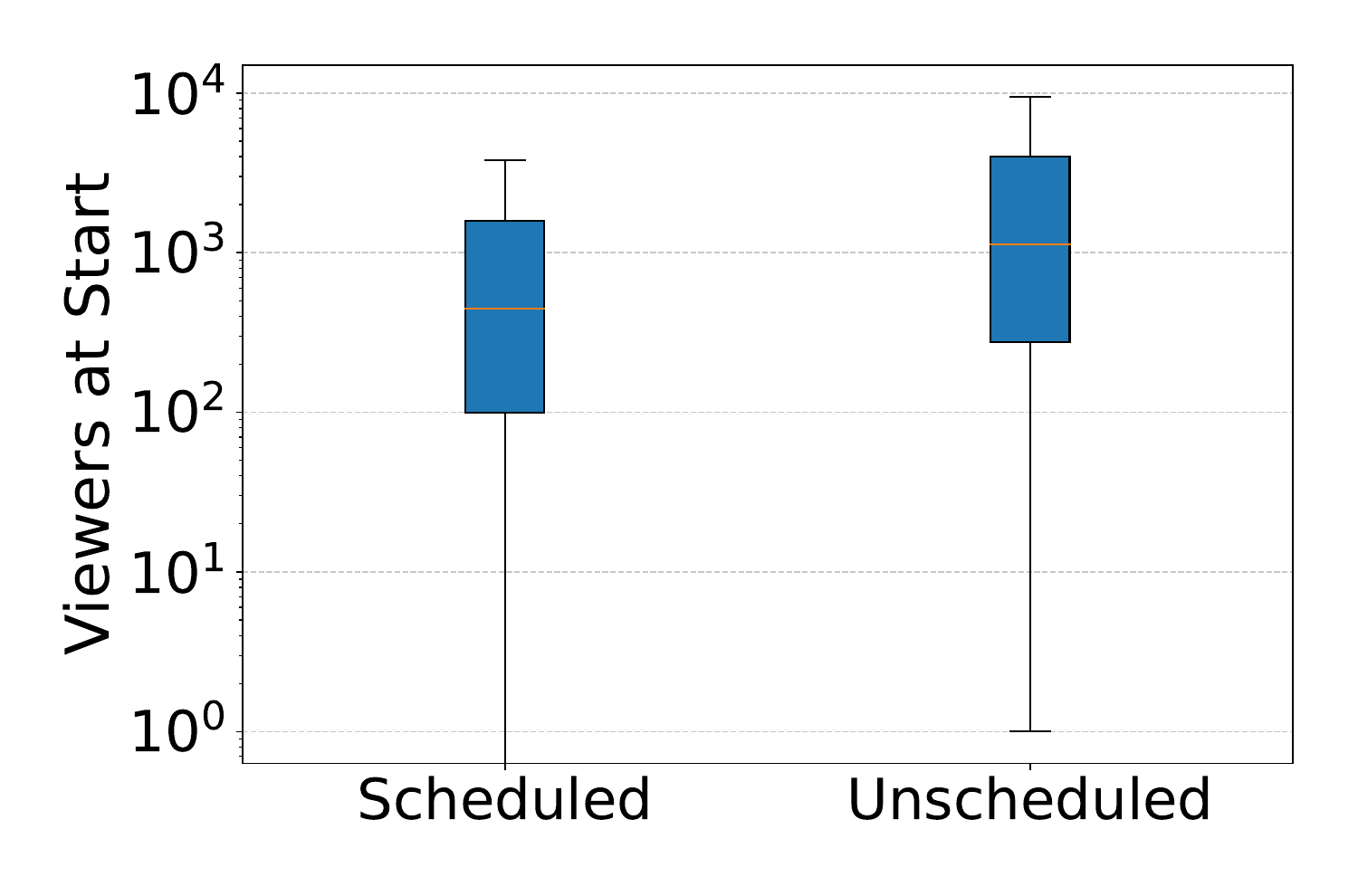}
        \caption{\small{Effect of scheduling on the initial number of viewers}}
        \label{fig:bp:stats:sched-vs-unsched}
    \end{minipage}    
        \vspace{-6mm}
\end{figure}

To provide deeper insight into the dataset, we group the videos into three categories based on their duration: short (under 120 minutes), medium (between 120 and 240 minutes), and long (more than 240 minutes). Figure~\ref{fig:categorized} shows the viewership patterns across these three categories. Specifically, Figure~\ref{fig:categorized:ubm} shows the average views for each video category on weekdays and weekends. Short videos get the highest views, medium ones are in the middle, and long videos get the least. This pattern is the same on both weekdays and weekends, but all of them get a bit more views on weekends. Figure~\ref{fig:categorized:ubsom} shows how stable the views are using CV. A lower ratio means the views are more consistent. Short videos are the most stable, while long ones are the most random. Medium videos are somewhere in between. So, short videos are not only more popular, but their performance is also more predictable. Long videos are less popular and their view counts change a lot. Figure~\ref{fig:categorized:cdfs} presents the viewer count CDFs for short, medium, and long videos, plotted together with separate x-axes for clarity. The blue curve (short videos) shows that streams shorter than $40$ minutes are relatively rare, while about $80\%$ are under $100$ minutes. The green curve (medium videos) rises steadily, indicating that their durations are spread fairly evenly between $120$ and $220$ minutes. The purple curve (long videos) rises quickly at first and then flattens, suggesting that extremely long streams are uncommon, with over $80\%$ shorter than $2000$ minutes (about $33$ hours). The figure also marks the first and third quartiles for each category. The inter-quartile range for short and long videos is narrower than for medium videos, highlighting that medium video durations are more variable.
\begin{figure}[t]
  \centering
  \begin{subfigure}[b]{0.45\linewidth}
    \centering
    \includegraphics[width=\linewidth]{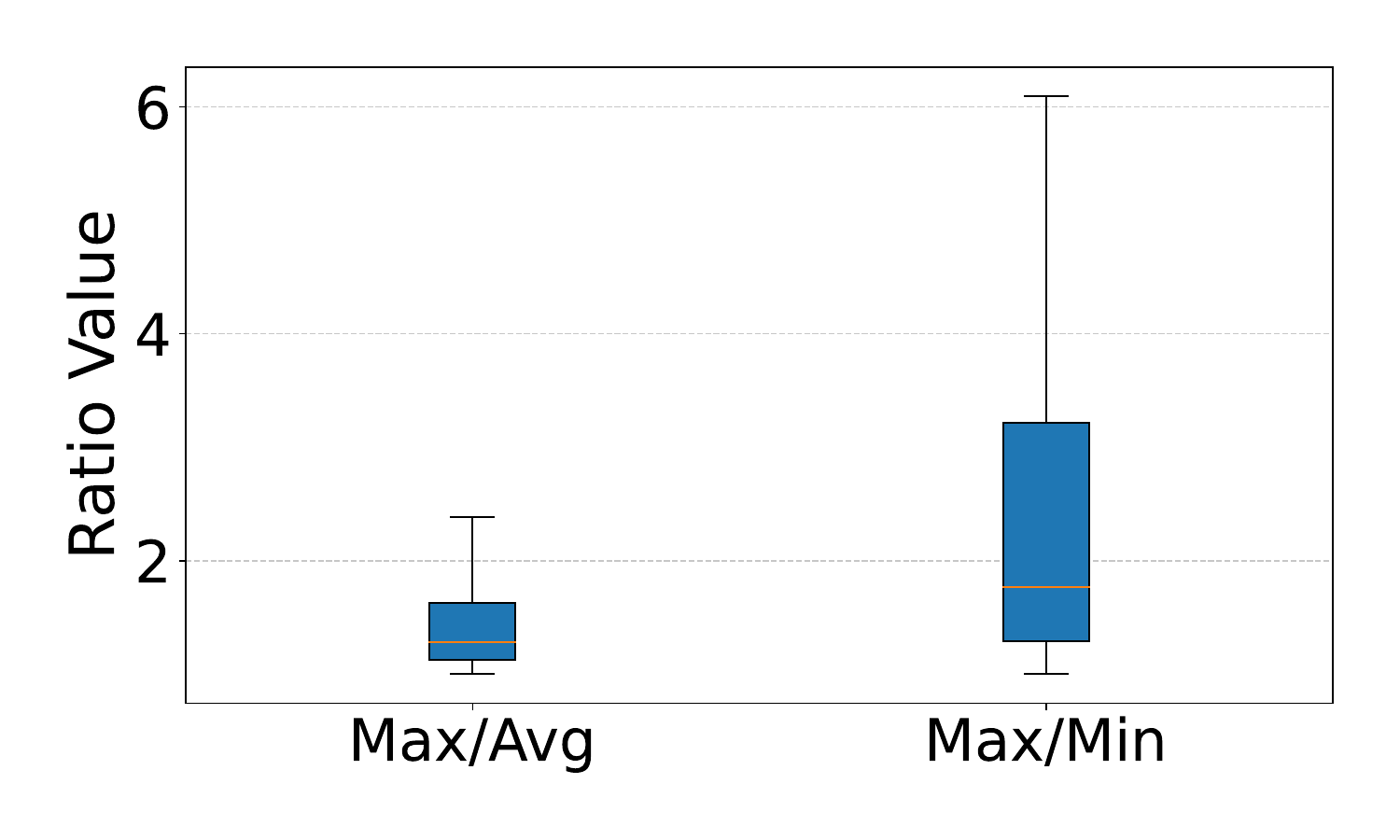}
    \caption{\small{Ratio of max. to min. and max. to avg. number of concurrent viewers} 
    }
    \label{fig:bp:stats:ration-value}
  \end{subfigure}
  \hfil
  \begin{subfigure}[b]{0.45\linewidth}
    \centering
    \includegraphics[width=\linewidth]{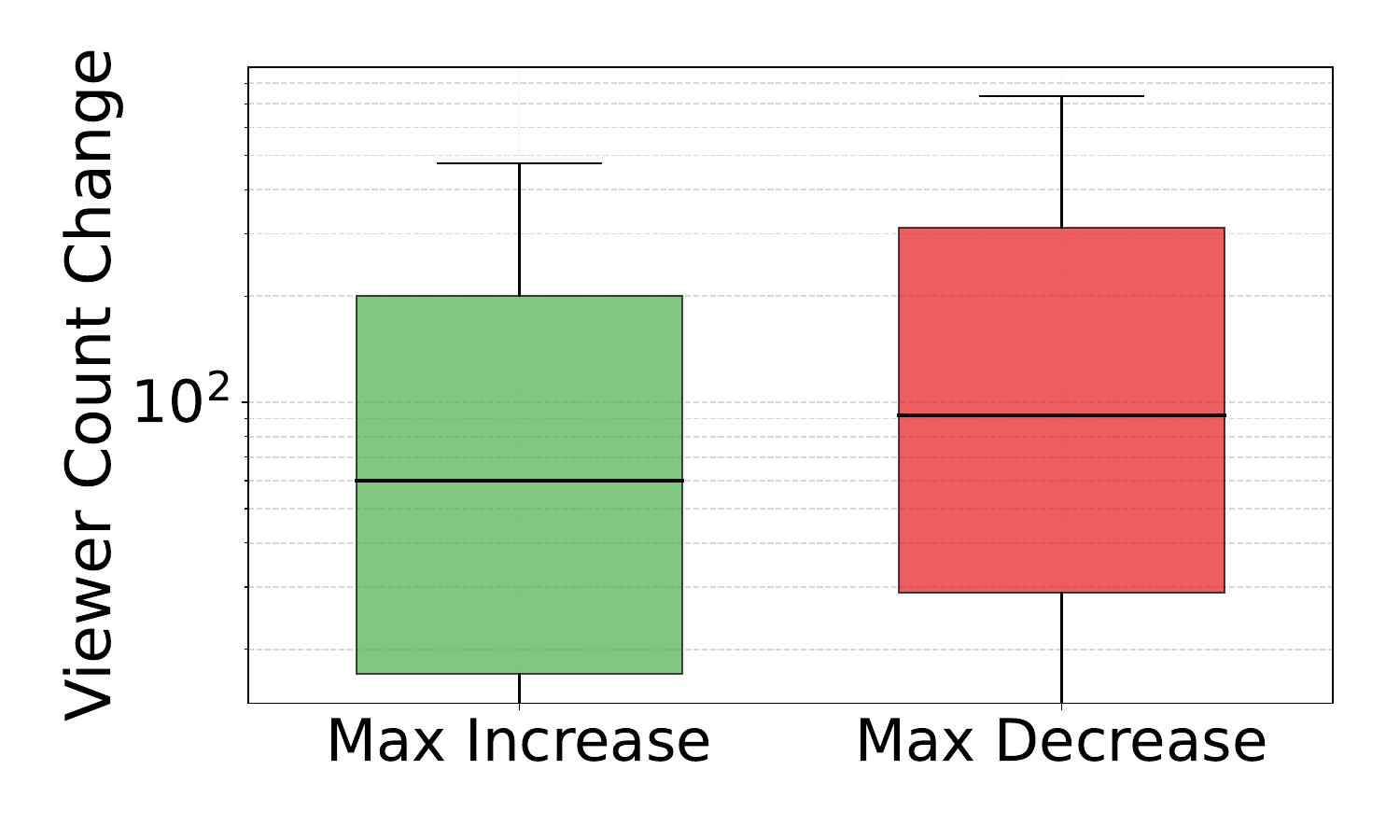}
    \caption{\small{Maximum consecutive increase and decrease in number of concurrent viewers}
    }
    \label{fig:bp:stats:consecutive-change}
  \end{subfigure}
  \caption{\small{Pattern of change in number of concurrent viewers}}
      \vspace{-4mm}
\end{figure}

YouTube streams can be either scheduled or unscheduled. Scheduling announces the start time to subscribers and may act as a form of advertisement. However, our data does not show a positive effect of scheduling. In fact, scheduled streams have a lower median number of viewers at the five-minute mark compared to unscheduled streams. It is important to note that scheduled streams are much more frequent in our dataset, which may influence this observation. Specifically, from about $12000$ streams in the dataset, only $1000$ are not scheduled. Further analysis is needed to better understand the effect of scheduling on the number of views at the start of the stream.

To analyze short-term fluctuations in viewership, we examine the variation within each stream by calculating the ratio of the maximum number of viewers to both the average and the minimum number of viewers recorded over $5$-minute intervals. The results, shown in Figure~\ref{fig:bp:stats:ration-value}, indicate that the third quartile of the maximum-to-average ratio is below $2$, suggesting moderate variation around the average. The maximum-to-minimum ratio is below $4$ for most streams. We also compute the maximum increase and maximum decrease in the number of viewers between consecutive $5$-minute intervals. As shown in Figure~\ref{fig:bp:stats:consecutive-change}, the median change ranges from $60$ to $90$ viewers, with the third quartile remaining below $300$.

\section{Application}\label{sec:application}
The YTLive dataset focuses on concurrent viewership and precise stream durations, offering a suitable resource for diverse research applications. It enables large-scale, quantitative analysis of audience dynamics and content temporality on YouTube Live.
For systems and infrastructure researchers, it provides realistic traces of viewership dynamics essential for designing and validating resource allocation schemes. Metrics like stream duration distributions and viewer volatility help build accurate workload models for testing CDN caching and server provisioning. The ALPHAS system~\cite{11044484} demonstrates a direct application, where an early version of this dataset helped define the problem of multi-stream resource optimization.

For data scientists and content analysts, the dataset supports behavioral research on YouTube Live. The viewer time-series enables the development of viewership prediction models, as explored by Chen \etal~\cite{10.1007/978-981-97-0837-6_3}. It also facilitates analysis of factors like the link between stream duration and audience growth~\cite{le2021studychannelpopularitytwitch}, and supports validation of findings on audience size and engagement~\cite{10.1177/14614448211069996}. While lacking fine-grained user interaction data, the dataset offers valuable context. For instance, it complements studies like Jia \etal~\cite{10.1145/2957750} by providing a macro-level view of audience scale.

\section{Conclusion}\label{sec:conclusion}
This paper presented YTLive, a large-scale and publicly available~\cite{YTLivedataset} dataset of YouTube Live streams collected over two-month period in May and June 2024. The dataset includes \liveCVrecords records from \livesession YouTube live sessions, capturing concurrent viewer counts every five minutes and detailed information on stream durations. We developed an automated, modular data collection system that ensures high-quality and timely data using the YouTube Data API~\cite{yda}.
Our initial analysis of the dataset showed clear patterns in live viewer behavior. For example, viewership tends to be higher and more stable on weekends and during afternoon hours. We also observed that shorter streams generally attract larger and more consistent audiences compared to longer ones. These findings can support future research in areas such as adaptive streaming, resource management, and QoE modeling.
By sharing YTLive as an open dataset, we aim to provide a valuable resource for the research community. We hope it will enable reproducible studies and help improve the design and performance of live streaming systems.



\balance
\bibliographystyle{bib/IEEEtran}
\bibliography{bib/ref}
\end{document}